# Bayesian Inference for Multidimensional Welfare Comparisons

**David Gunawan**

School of Mathematics and Applied Statistics, University of Wollongong, Wollongong, NSW 2522, Australia (dgunawan@uow.edu.au)

**William Griffiths***

Department of Economics, University of Melbourne, Melbourne, Vic 3010, Australia (wegrif@unimelb.edu.au)

**Duangkamon Chotikapanich**

Department of Econometrics and Business Statistics, Monash University, Melbourne, Vic 3145, Australia (duangkamon.chotikapanich@monash.edu)

## Abstract

Using both single-index measures and stochastic dominance concepts, we show how Bayesian inference can be used to make multivariate welfare comparisons. A four-dimensional distribution for the well-being attributes income, mental health, education, and happiness are estimated via Bayesian Markov chain Monte Carlo using unit-record data taken from the Household, Income and Labour Dynamics in Australia survey. Marginal distributions of beta and gamma mixtures and discrete ordinal distributions are combined using a copula. Improvements in both well-being generally and poverty magnitude are assessed using posterior means of single-index measures and posterior probabilities of stochastic dominance. The conditions for stochastic dominance depend on the class of utility functions that is assumed to define a social welfare function and the number of attributes in the utility function. Three classes of utility functions are considered, and posterior probabilities of dominance are computed for one, two, and four-attribute utility functions for three time intervals within the period 2001 to 2019.

# 1. INTRODUCTION

It is well recognised that social welfare is multidimensional, including many well-being attributes such as levels of income, health, and education. Inadequate levels of well-being attributes (multidimensional poverty) and large variations in individual well-being attributes (multidimensional inequality) contribute negatively to social welfare. Influenced by a pioneering paper by Kolm (1977), multivariate approaches to well-being developed along two lines, analogous to univariate approaches to inequality and poverty measurement. The first approach is concerned with deriving indices of multidimensional inequality and poverty. For examples of inequality indices see Maasoumi and Zandvakili (1986) and Lugo (2007); for examples of deprivation/poverty indices see Bourguignon and Chakravarty (2003), and Alkire and Foster (2011). Widely used measures at national and international levels are the Human Development Index (HDI) and the Human Poverty Index (HPI) used by the United Nations Development Program. These indices measure a country's average achievements in three basic aspects of human development: a long and healthy life, access to knowledge (education), and income. An advantage of the multidimensional index approach is that it provides a complete ordering of the joint distributions of well-being attributes when comparisons are being made. A disadvantage is that it involves choosing two aggregation functions, one over the individuals and the other over well-being attributes, and choosing arbitrary key parameters, such as the relative weights for each well-being distribution and substitution parameters between well-being distributions. This problem was partially alleviated in a proposal by Maasoumi and Racine (2016). In their approach, the weights and substitution parameters for the multidimensional inequality index proposed by Maasoumi (1986) can be varied for different quantiles of the bivariate well-being distribution, and are estimated using a nonparametric method. This approach provides a robust solution for assigning weights and substitution parameters, but the loss of information from considering a single index rather than a joint well-being distribution remains.

The second approach compares the joint distributions of well-being attributes on the basis of derived multidimensional stochastic dominance conditions. Examples are Atkinson and Bourguignon (1982), Duclos et al. (2006), and Muller and Trannoy (2011). This approach allows for agreement over broad classes of social welfare functions and over different forms of aggregating well-being attributes without any precise knowledge about the underlying social welfare function, and avoids choosing particular aggregation functions and arbitrary key parameters. However, at best it can only provide a partial ordering of distributions. When the dominance conditions do not hold, results from comparisons of multivariate distributions are inconclusive. Atkinson and Bourguignon (1982) derived bivariate welfare dominance conditions for various classes of utility functions defined by the signs of their derivatives up to fourth order. Their framework was extended by Muller and Trannoy (2011, 2012) who proposed bivariate and trivariate welfare dominance conditions which allow for an asymmetric treatment of different welfare measures. For example, a high income can be used to compensate for bad health or a low level of education. The conditions in Muller and Trannoy (2011, 2012) are extended in this paper by providing four-variate dominance conditions. Duclos et al. (2006) and Bourguignon and Chakravarty (2003) also adapted the Atkinson and Bourguignon (1982) framework to study bivariate poverty comparisons.

Comparisons of multivariate well-being distributions using inequality and poverty measures or stochastic-dominance relationships typically use samples of well-being attributes from survey datasets. Such comparisons do not lead to a conclusion that can be made with certainty because they are subject to statistical sampling error. This uncertainty has led to the development of a number of statistical tests for univariate and multivariate stochastic dominance, tests that can be used to compare distributions at two different points in time or for two different population subgroups. An extensive list of studies using univariate stochastic dominance tests is given in Lander et al. (2020). Univariate stochastic dominance tests suggested by Anderson (1996) and Davidson and Duclos (2000) were extended to multivariate frameworks by Crawford (2005) and Duclos et al. (2006), respectively. McCaig and Yatchew (2007) propose univariate and multivariate stochastic dominance tests that are based on a test proposed by Hall and Yatchew (2005).

In contrast to these frequentist testing approaches, we extend the Bayesian approach proposed by Lander et al. (2020) and assess whether one multivariate distribution stochastically dominates another by computing the posterior probability of dominance. Three probabilities can be computed: the probability that a distribution $X$ dominates another, $Y$, the probability $Y$ dominates $X$, and the probability there is no dominance. This approach is not a formal statistical testing approach that requires the setting of a null hypothesis such as dominance or no dominance, but a means of providing expost sample information on the probability of each of the three "states of nature". Reporting dominance probabilities for the three possible outcomes is more informative than the frequentist sampling-theory procedure of reporting rejection or non-rejection of the null hypothesis. An extensive discussion on the differences between frequentist and Bayesian approaches is given in Lander et al (2020). Using the Bayesian approach, it is also possible to focus on some segments of the population, like multidimensionally poor people. Furthermore, we can make a probability statement about the values of welfare, inequality, and poverty measures, such as the multidimensional poverty measures proposed by Alkire and Foster (2011). See Section 6 for further details.

Our application contributes to the multidimensional well-being and poverty literature in Australia over the period 2001 – 2019 using data from the Household, Income, and Labour Dynamics in Australia (HILDA) survey (Watson and Wooden, 2012). Four essential well-being attributes are considered: income, mental health, education and happiness. Observations on income and mental health are such that they can be conveniently modelled using continuous distributions, a mixture of gamma densities for income and a mixture of beta distributions for mental health. Observations on education and happiness are such that they are modelled as ordinal categorical variables. These marginal distributions are combined into a four-dimensional multivariate distribution using a copula. The parameters of the multivariate distribution are estimated using Bayesian Markov chain Monte Carlo (MCMC). The posterior probabilities are then computed as the proportion of MCMC draws that satisfy the conditions that define the dominance criteria.

Several stochastic dominance conditions are considered. These conditions depend on the class of utility functions used to define a social welfare function, and the number of well-being attributes in the utility function. We consider three classes of utility functions, with each class defined by the signs of its derivatives, and the resulting stochastic dominance conditions when utility is a function of one, two, or four attributes. Results are presented for (a) the marginal distributions of income, mental health, education, and happiness, relevant for single-attribute utility functions, (b) bivariate distributions for income with each of the other attributes, relevant for two-attribute utility functions, and (c) the four-dimensional distribution for all attributes, relevant when they are all included in the utility function.

In Section 2, we describe the data on each of the attributes and the marginal distributions used to model each of them. Section 3 contains details of how the marginal distributions are combined to form a multivariate distribution using a copula. Estimation is discussed in Section 4, with the details of the MCMC algorithm provided in the supplementary material. Changes in well-being and poverty over three time intervals are considered in Sections 5 and 6, respectively. The time intervals are 2001 to 2010, 2010 to 2015 and 2015 to 2019. In Section 5, we describe the three classes of utility functions, their implications for stochastic dominance conditions, and a single index measure that can be used to assess whether well-being has improved. In Section 6 we give results for a single index measure of poverty and a restricted form of stochastic dominance that can be used to assess whether poverty outcomes have improved. Some concluding remarks are given in Section 7.

## 2. DATA AND MARGINAL DISTRIBUTIONS

Data were obtained from four waves of the HILDA survey corresponding to years 2001, 2010, 2015, and 2019. These years were chosen to examine a variety of pairwise multidimensional stochastic comparisons and to track improvements (or deteriorations) in well-being distributions over time. The distributions for each of the years are estimated independently. Cross-sectional weights provided by the HILDA dataset are used to correct the representativeness of the HILDA sample to the population.

The variables chosen for welfare comparisons were (1) household disposable income converted to a per individual basis, (2) the SF-36 mental health score, (3) education level, and (4) happiness. The observational units were all individuals age 15 and above; those aged less than 15 were omitted because of the unavailability of the SF-36 mental health score, education and happiness levels and sampling weights for these individuals. However, the total number of children was still used in the calculation of equivalised income.

### 2.1 Income

To obtain net disposable income we subtracted "total disposable income negative per household" from "total disposable income positive per household". The conversion of this variable to a per individual basis involves choice of an equivalence scale and the allocation of equivalised income to members of each household. For these steps we followed Sila and Dugain (2019). To obtain equivalised income, net disposable income was divided by the square root of the number of individuals in the household. This quantity was assigned to each member of the household. Values were deflated using the Consumer Price Index, treating 2000/2001 as the base. Incomes that were non-positive were omitted from the samples.

For modelling incomes that are only defined on the positive real line $(0,\infty)$, a mixture of gamma densities is a good candidate: it can exhibit skewness, multi-modality, and heavy tails, possible properties of an income distribution. Denoting income by $Y_1$, an income distribution that follows a gamma mixture with $K_G$ components can be written as:

$$p(y_1 \mid \xi_G, \mu, \nu) = \sum_{k=1}^{K_G} \xi_{k,G} G(y_1 \mid \nu_k, \mu_k)$$

where $y_1$ is a random draw from the probability density function (pdf), $p(y_1 \mid \xi_G, \mu, \boldsymbol{v})$, with parameter vectors, $\xi_G = (\xi_{1,G}, \xi_{2,G}, \ldots, \xi_{K_G,G})'$, $\mu = (\mu_1, \mu_2, \ldots, \mu_{K_G})'$, and $\boldsymbol{v} = (v_1, v_2, \ldots, v_{K_G})'$. The pdf $G(y_1 \mid v_k, \mu_k)$ is a gamma density with mean $\mu_k > 0$ and shape parameter $v_k > 0$. That is,

$$G(y_1 \mid v_k, \mu_k) = \frac{(v_k/\mu_k)^{v_k}}{\Gamma(v_k)} y_1^{v_k-1} \exp\left\{\frac{v_k y_1}{\mu_k}\right\}$$

### 2.2 Mental Health

The mental health score is extracted from responses to the SF-36 health survey questions provided in the HILDA survey. The SF-36 survey is a multipurpose and short form health survey with 36 questions that provide one of the most widely used generic and continuous measures of health-related quality of life in clinical research and general population health. It has been translated and studied in more than 40 countries (Ware et al., 1993). The developers of the SF-36 claim that scaling assumptions used to transform the ordered categorical responses into a continuous health measure can be interpreted as "quasi-interval measurement scales" (Ware and Gandek, 1998). They argue that such scales can consistently rank health status, and that the ratio of differences between scores has meaning. While such claims can be debatable, it is important to note that all techniques used to convert discrete category scores into a continuous variable will have some issues. Evidence provided by Butterworth and Crosier (2004) supports the validity of SF-36 data collected by the HILDA survey as general measures of physical and mental health status. Scales for both physical and mental health are provided. We are concerned with the mental health scale. Details of how it is constructed can be found in Ware et al. (1993). The final measure ranges between 0 and 100, where a score of 100 implies good health and a 0 represents a serious mental health problem. Individuals with scores below 50 are considered to have poor mental health. We normalized the score such that it lies between 0 and 1.

A mixture of beta distributions is employed to model the mental health scores, denoted by $Y_2$, on the $[0,1]$ interval. The most common parameterization of a beta density in terms of parameters $\alpha_k$ and $\beta_k$ is

$$B(y_2 \mid \alpha_k, \beta_k) = \frac{\Gamma(\alpha_k + \beta_k)}{\Gamma(\alpha_k)\Gamma(\beta_k)} y_2^{\alpha_k - 1}(1 - y_2)^{\beta_k - 1}$$

We employ a different parameterization such that the mean $m_k = \alpha_k / (\alpha_k + \beta_k)$ is one of the parameters. This parameterization is given by:

$$B(y_2 \mid s_k, m_k) = \frac{\Gamma(s_k)}{\Gamma(s_k m_k)\Gamma(s_k(1 - m_k))} y_2^{s_k m_k - 1}(1 - y_2)^{s_k(1-m_k)-1}$$

where $s_k = \alpha_k + \beta_k$. The mixture of beta densities with $K_B$ components is given by:

$$p(y_2 \mid \boldsymbol{s}, \boldsymbol{m}, \xi_B) = \sum_{k=1}^{K_B} \xi_{k,B} B(y_2 \mid s_k, m_k)$$

where $y_2$ is a random draw of a mental health score from the probability density function (pdf) $p(y_2 \mid \boldsymbol{s}, \boldsymbol{m}, \xi_B)$ with parameter vectors $\xi_B = (\xi_{1,B}, \xi_{2,B}, \ldots, \xi_{K_B,B})'$, $\boldsymbol{s} = (s_1, s_2, \ldots, s_{K_B})'$ and $\boldsymbol{m} = (m_1, m_2, \ldots, m_{K_B})'$.

### 2.3 Education and Happiness

The variables EDHIGH1 and LOSAT, describing the highest level of education achieved and the happiness level, respectively, were extracted from the HILDA survey. Both variables are ordered categorical variables with 10 categories; in each case we condensed those 10 categories to 5. The 5 categories for education $(y_3)$ are given in Table 1. The LOSAT variable ranges from totally dissatisfied

(0) to totally satisfied (10); the way in which these scores were condensed to obtain the happiness variable $(y_4)$ are displayed in Table 2.

Table 1. Categories for Highest Education Level Achieved

| Category | Explanation |
|---|---|
| 1 | Year 11 or below |
| 2 | Year 12 |
| 3 | Cert I, II, III or IV, and Certificate not defined |
| 4 | Advanced Diploma, Diploma, Bachelor or Honours, Graduate Diploma, and Graduate Certificate |
| 5 | Postgraduate – Masters or Doctorate |

Table 2. Categories for Happiness

| Category | Explanation |
|---|---|
| 1 | Scores 0 to 2 are combined |
| 2 | Scores 3 to 4 are combined |
| 3 | Scores 5 to 6 are combined |
| 4 | Scores 7 to 8 are combined |
| 5 | Scores 9 to 10 are combined |

Standard ordinal data models are employed for the education and happiness variables. The models for these variables can be thought of as arising from an underlying latent variable threshold-crossing framework. This framework can be motivated by assuming a continuous latent variable $Y^*$, where $Y^* \sim N(0,1)$. The outcome variables $Y_3$ and $Y_4$ arise according to the following set up:

$$Y_j = s, \qquad \text{if } \tau_{j,s-1} < Y_j^* \leq \tau_{j,s}$$

where $-\infty = \tau_{j,0} < \tau_{j,1} < \ldots < \tau_{j,L_j} = \infty,$ for $j = 3,4,$ are threshold parameters that need to be estimated and $L_j$ is the number of categories of the $j$-th variable. The threshold parameters determine the discretization of the education and happiness data into the $L_3$ and $L_4$ ordered categories, respectively. The proportion of population that belongs to the $s$-th level, $p_{j,s}$ can be written as $\Phi(\tau_{j,s}) - \Phi(\tau_{j,s-1})$ for $j = 3,4$, where $\Phi(\cdot)$ denotes the standard normal distribution function.

## 3. MULTIVARIATE DISTRIBUTION

A convenient way to construct a multivariate distribution is via a copula. To introduce the copula, we denote the distribution functions for each of the attributes by $\tilde{U}_j = F_j\left(y_j \mid \boldsymbol{\theta}_j\right), j = 1,2,3,4,$ where $\boldsymbol{\theta}_j$ is a vector of parameters relating to the distribution of the $j$-attribute. That is $\boldsymbol{\theta}_1' = \left(\boldsymbol{\xi}_G' , \boldsymbol{\mu}', \boldsymbol{v}'\right)$, $\boldsymbol{\theta}_2' = \left(\boldsymbol{\xi}_B' , \boldsymbol{s}', \boldsymbol{m}'\right)$, and $\boldsymbol{\theta}_j' = \left(\tau_{j,1}, \tau_{j,2}, \ldots, \tau_{j,L_j-1}\right)$ for $j = 3,4$. Let $\boldsymbol{\Theta} = \{\boldsymbol{\theta}_1, \boldsymbol{\theta}_2, \boldsymbol{\theta}_3, \boldsymbol{\theta}_4\}$ be the complete set of parameters in the marginal distributions. Following Sklar (1959), a joint distribution function $F$ with marginal distribution functions $F_1(y_1 \mid \theta_1)$, $F_2(y_2 \mid \theta_2)$, $F_3(y_3 \mid \theta_3)$, $F_4(y_4 \mid \theta_4)$ can be written as

$$F\left(y_1, y_2, y_3, y_4 \mid \boldsymbol{\Theta}, \boldsymbol{\Gamma}\right) = C\left(F_1(y_1 \mid \boldsymbol{\theta}_1), F_2(y_2 \mid \boldsymbol{\theta}_2), F_3(y_3 \mid \boldsymbol{\theta}_3), F_4(y_4 \mid \boldsymbol{\theta}_4) \mid \boldsymbol{\Gamma}\right)$$

where $C$ is a copula distribution function and $\boldsymbol{\Gamma}$ is a set of parameters in the copula function. The function $C$ is a multivariate joint distribution function defined on the 4-dimentional $[0,1]$ region with uniform distributions on the interval $[0,1]$ for the marginal distributions for $\tilde{U}_1, \tilde{U}_2, \tilde{U}_3, \tilde{U}_4$. More details about copulas can be found in Trivedi and Zimmer (2005) and Nelsen (2006). In this paper, we use the popular Gaussian copula model (Song, 2000).

In the Gaussian copula, each $\tilde{U}_j$ is transformed to a standard normal random variable $Y_j^* = \Phi^{-1}(\tilde{U}_j)$ such that the correlation matrix for $\left(Y_1^*, Y_2^*, Y_3^*, Y_4^*\right)$ is $\boldsymbol{\Gamma}$. For the continuous random variables ($j = 1$ and 2) this transformation is one to one, and hence straightforward; $Y_j^* = \Phi^{-1}\left(F_j(y_j \mid \boldsymbol{\theta}_j)\right)$. However, when the marginal distribution is ordinal-valued as in the case of the education and happiness distributions, the transformations $Y_3 \to \tilde{U}_3$ and $\tilde{U}_3 \to Y_3^*$ and $Y_4 \to \tilde{U}_4$ and $\tilde{U}_4 \to Y_4^*$ are both one to many. Pitt et al. (2006) point out that if the *j*th margin is discrete or ordinal-valued, the problem of estimation can be greatly simplified by treating $Y_j^*$ as latent variables. For the Gaussian copula the latent variables $Y_3^*$ and $Y_4^*$ can be generated explicitly from truncated Gaussian distributions in an MCMC scheme. For $j=3,4$, the bounds on the latent variable $Y_j^*=\Phi^{-1}(\tilde{U}_j)$ are given by:

$$\Phi^{-1}\left(F_j(y_j-1)\,|\,\theta_j\right)\le Y_j^*<\Phi^{-1}\left(F_j(y_j)\,|\,\theta_j\right)$$

This scheme is discussed in Section B of the online supplementary material. The Gaussian copula distribution is a multivariate normal distribution defined over $\left(Y_1^*,Y_2^*,Y_3^*,Y_4^*\right)$. Its distribution and density functions are given by (Song, 2000):

$$C\left(\tilde{u}_1,\tilde{u}_2,\tilde{u}_3,\tilde{u}_4\,|\,\boldsymbol{\Gamma},\boldsymbol{\Theta}\right)=\Phi_G\left(y_1^*,y_2^*,y_3^*,y_4^*\,|\,\boldsymbol{\Gamma},\boldsymbol{\Theta}\right)$$

and

$$\begin{aligned}c\left(\tilde{u}_1,\tilde{u}_2,\tilde{u}_3,\tilde{u}_4\,|\,\boldsymbol{\Gamma},\boldsymbol{\Theta}\right)&=\frac{\partial C\left(\tilde{u}_1,\tilde{u}_2,\tilde{u}_3,\tilde{u}_4\,|\,\boldsymbol{\Gamma},\boldsymbol{\Theta}\right)}{\partial\tilde{u}_1\partial\tilde{u}_2\partial\tilde{u}_3\partial\tilde{u}_4}\\&\propto|\Gamma|^{-1/2}\exp\left\{-\frac{1}{2}\boldsymbol{y}^{*\prime}\left(\boldsymbol{\Gamma}^{-1}-\boldsymbol{I}\right)\boldsymbol{y}^*\right\}\end{aligned}\tag{1}$$

where $\boldsymbol{y}^*=\left(y_1^*,y_2^*,y_3^*,y_4^*\right)'$ and $\Phi_G(\cdot)$ is the distribution function of the standard 4-dimensional multivariate Gaussian distribution $N(0,\Gamma)$.

## 4. ESTIMATION

There are two sets of parameters that require estimation. The first set is the parameters of each of the selected marginal distributions, $\boldsymbol{\Theta}=\left\{\boldsymbol{\theta}_1,\boldsymbol{\theta}_2,\boldsymbol{\theta}_3,\boldsymbol{\theta}_4\right\}$, where $\boldsymbol{\theta}_1=\left\{\xi_G,\mu,\boldsymbol{v}\right\}$, $\boldsymbol{\theta}_2=\left\{\xi_B,\boldsymbol{s},\boldsymbol{m}\right\}$, and $\boldsymbol{\theta}_j=(\tau_{j,1},\tau_{j,2},\ldots,\tau_{j,L_j-1})'$ for $j=3$ and 4. The second set is the dependent parameters of the Gaussian copula model $\Gamma$.

We consider an independent sample of $n$ observations, $\boldsymbol{y}'=(\boldsymbol{y}_1',\boldsymbol{y}_2',\ldots,\boldsymbol{y}_n')$, where $\boldsymbol{y}_i'=\left(y_{i,1},y_{i,2},y_{i,3},y_{i,4}\right)$. To estimate $\boldsymbol{\Theta}=\left\{\boldsymbol{\theta}_1,\boldsymbol{\theta}_2,\boldsymbol{\theta}_3,\boldsymbol{\theta}_4\right\}$ and $\boldsymbol{\Gamma}$, we augment the likelihood function with the transformed Gaussian copula variables $\boldsymbol{y}_i^*=\left(y_{i,1}^*,y_{i,2}^*,y_{i,3}^*,y_{i,4}^*\right)'$, for $i=1,2,\ldots,n.$ The augmented likelihood is given by

$$
\begin{aligned}
p\left(\boldsymbol{y},\boldsymbol{y}^{*}\mid\boldsymbol{\Theta},\boldsymbol{\Gamma}\right) &= \prod_{i=1}^{n} p\left(\boldsymbol{y}_i,\boldsymbol{y}_i^{*}\mid\boldsymbol{\Theta},\boldsymbol{\Gamma}\right) \\
&= \prod_{i=1}^{n} p\left(\boldsymbol{y}_i^{*}\mid\boldsymbol{\Theta},\boldsymbol{\Gamma}\right) p\left(\boldsymbol{y}_i\mid\boldsymbol{y}_i^{*},\boldsymbol{\Theta},\boldsymbol{\Gamma}\right) \\
&= \prod_{i=1}^{n} c\left(y_{i,1}^{*},y_{i,2}^{*},y_{i,3}^{*},y_{i,4}^{*}\mid\boldsymbol{\Theta},\boldsymbol{\Gamma}\right) I\left(a_{i,3}\le y_{i,3}^{*}<b_{i,3}\right) I\left(a_{i,4}\le y_{i,4}^{*}<b_{i,4}\right)\prod_{j=1}^{2} p_j\left(y_{i,j}\mid\boldsymbol{\theta}_j\right)
\end{aligned}
$$

where $y_{i,j}^{*}=\Phi^{-1}\left(F_j(y_{ij})\mid\boldsymbol{\theta}_j\right)$ for $j=1,2,$ $a_{i,j}=\Phi^{-1}\left(F_j(y_{i,j}-1)\mid\boldsymbol{\theta}_j\right)$ and $b_{i,j}=\Phi^{-1}\left(F_j(y_{i,j})\mid\boldsymbol{\theta}_j\right)$ for $j=3,4,$ and $c\left(y_{i,1}^{*},y_{i,2}^{*},y_{i,3}^{*},y_{i,4}^{*}\mid\boldsymbol{\Theta},\boldsymbol{\Gamma}\right)$ is given in Equation (1). We assume the prior density is of the form:

$$
p(\boldsymbol{\Gamma},\boldsymbol{\Theta}) = p(\boldsymbol{\Gamma})\prod_{j=1}^{4} p\left(\boldsymbol{\theta}_j\right)
$$

The prior distributions that we specify for all the unknown parameters are relatively uninformative. Details are provided in Section A of the supplementary material.

The posterior density is given by

$$
p\left(\boldsymbol{y}^{*},\boldsymbol{\Theta},\boldsymbol{\Gamma}\mid\boldsymbol{y}\right) \propto p\left(\boldsymbol{y}^{*},\boldsymbol{y}\mid\boldsymbol{\Theta},\boldsymbol{\Gamma}\right) p(\boldsymbol{\Theta},\boldsymbol{\Gamma})
$$

Details of an MCMC algorithm for generating observations on the parameters from the posterior density are given in Section B of the online supplementary material. This algorithm combines that for estimating the multivariate well-being model with one that accommodates the cross-sectional sampling weights provided with the HILDA data. A Bayesian bootstrap algorithm (Gunawan et al., 2020b) is used to generate pseudo representative random samples to correct the representativeness of the HILDA sample to the population. The number of components in the mixture models was chosen by estimating distribution functions for 1, 2, 3 and 4 components and comparing their estimated distribution function values with the empirical distribution functions of hold-out samples. The chosen values of $K_G$ and $K_B$ were those that minimised the mean absolute error. A complete description of this process appears in Section C of the supplementary material. It led to gamma mixture models for income with two components in 2010, 2015 and 2019 and a model with three components in 2001. For the mental health score, three component beta mixture models were selected for all years. Having chosen the number of mixture components, the complete model was estimated using the full sample and the posterior draws $\{\boldsymbol{\Theta}^{(m)},\boldsymbol{\Gamma}^{(m)}\}_{m=1}^{M}$ were used to compute posterior probabilities of dominance and summary statistics for the posterior densities of single index measures. Plots of the predictive densities for each of the attributes are displayed in Section D of the supplementary material.

# 5. WELL-BEING

To examine changes in well-being, we consider both single index measures and probabilities of dominance. When considering each attribute separately we use the mean as a single index measure and univariate first and second order stochastic dominance. Bivariate comparisons involving income and each of the other attributes, as well as a 4-dimensional comparison using all attributes are made with a multivariate welfare index (MWI) as a single index measure and with bivariate or multivariate stochastic dominance.

In what follows we introduce the general framework for examining stochastic dominance, then, in subsections defined by specific assumptions about the utility function that defines social welfare, we present results for both single index measures and the stochastic dominance concepts relevant for each utility function.

## 5.1 General Framework

We denote the joint density and cumulative distribution functions (cdfs) for the four attributes as $p(\boldsymbol{y})$ and $F(\boldsymbol{y})$, respectively, where $\boldsymbol{Y}=(Y_1,Y_2,Y_3,Y_4)$; $Y_1$ is income, $Y_2$ is the mental health score, $Y_3$ is the level of education, and $Y_4$ is happiness.[1] For deriving the multidimensional stochastic dominance conditions, we follow Muller and Trannoy (2011, 2012) and assume that the well-being density function $p(\boldsymbol{y})$ is defined on a finite support $[0,a_1]\times[0,a_2]\times[0,a_3]\times[0,a_4] = A_1\times A_2\times A_3\times A_4$, where $a_1,a_2,a_3,a_4$ are in $R^+$. This implies that each variable has a cardinal meaning, which is not satisfied by categorical education and happiness variables. This obstacle is overcome by assuming the latent continuous variables for education and happiness $Y_3^*$ for $Y_3$ and $Y_4^*$ for $Y_4$, that are defined on $A_3$ and $A_4$, respectively, are satisfactory. As in Muller and Trannoy (2012), we assume that, for the latent models with continuous distributions for $Y_3^*$ and $Y_4^*$, the utility function $U\left(y_1, y_2, y_3^*, y_4^*\right)$ is differentiable to the required degree with respect to any realized levels of $y_3^*$ of $Y_3^*$ and $y_4^*$ of $Y_4^*$. The additional assumption is that $U\left(y_1, y_2, y_3, y_4\right) = U\left(y_1, y_2, y_3^*, y_4^*\right)$. A social welfare function $W$ is defined as:

$$W = \int_0^{a_1} \int_0^{a_2} \int_0^{a_3} \int_0^{a_4} U\left(y_1, y_2, y_3, y_4\right) p\left(y_1, y_2, y_3, y_4\right) dy_4 \, dy_3 \, dy_2 \, dy_1$$

[1] The joint density and distribution functions depend on parameters, but we omit the dependence in this section for simplicity.

The change in social welfare between any two distribution functions $F_A(\boldsymbol{y})$ and $F_B(\boldsymbol{y})$ is given by:

$$\Delta W = W_A - W_B = \int\limits_{A_1 \times A_2 \times A_3 \times A_4} U\left(y_1, y_2, y_3, y_4\right) d\Delta F\left(y_1, y_2, y_3, y_4\right)$$

where $\Delta F$ denotes $F_A(\boldsymbol{y}) - F_B(\boldsymbol{y})$. Social welfare dominance is defined as unanimity over a family of social welfare functions defined by a given set of conditions on the partial derivatives of the utility functions. The social welfare in $A$ is no lower than in $B$ ($A$ dominates $B$) for a family $\boldsymbol{U}$ of utility functions if and only if the change in social welfare between $A$ and $B$ is nonnegative, $\Delta W = W_A - W_B \geq 0$, over the range of all possible values of $\boldsymbol{y}$.

For a given set of conditions on the utility function $U\left(y_1, y_2, y_3, y_4\right)$, social welfare dominance can be expressed in terms of sets of conditions on $\Delta F$ or on other functions of the distribution of $\boldsymbol{Y}$. In empirical work these functions are estimated, and, in Bayesian estimation that utilizes MCMC, each MCMC parameter draw yields a different estimate of functions for the distribution of $\boldsymbol{Y}$. Following Lander et al. (2020) and Gunawan et al. (2023), the posterior probability of dominance can be estimated as the proportion of MCMC draws that yield estimates of functions for the distribution of $\boldsymbol{Y}$ that satisfy the dominance conditions. Because income and mental health are continuous variables, when calculating the proportion of MCMC draws that satisfy the relevant conditions for all $y_1$ and $y_2$, the best we can do is to check the conditions for a finite grid of points. For income, we use 99 points between 5,495 and 150,000 with the intervals between successive values of $\ln(y_1)$ equal. The values 5495 and 150,000 are small enough and large enough to cover the minimum income and the maximum income, respectively in all of the 4 years. For mental health, we use 99 values from 0.01 to 0.99 in increments of 0.01. We use the five categorical values for education level and happiness. Precise expressions for the calculation of the dominance probabilities for three classes of utility functions are provided in Sections F and G of the supplementary material.

The three classes of utility functions that we consider are extensions of those proposed by Muller and Trannoy to four-dimensional well-being attributes. These classes are labelled as $\mathbf{U}^1$, $\mathbf{U}^2$ and $\mathbf{U}^3$. They are defined by their derivatives up to the 4th order.

### 5.2 Class of functions $\mathbf{U}^1$

Using subscripts to denote derivatives with respect to each of the attributes, the first and most general class $\mathbf{U}^1$ has the following signs on its derivatives.

$$\mathbf{U}^1 = \left\{ \begin{array}{c} U_1, U_2, U_3, U_4 \geq 0 \\ U_{12} \leq 0, U_{13} \leq 0, U_{14} \leq 0, U_{23} \leq 0, U_{24} \leq 0, U_{34} \leq 0 \\ U_{123} \geq 0, U_{124} \geq 0, U_{134} \geq 0, U_{234} \geq 0 \\ U_{1234} \leq 0 \end{array} \right\}$$

Increasing monotonicity of the utility function with respect to each attribute is implied by the assumption $U_1, U_2, U_3, U_4 \geq 0;$ well-being increases as each individual attribute increases. If utility is a function of only a single attribute, say $y_j$, the only relevant condition in $\mathbf{U}^1$ is $U_j \geq 0$ and the condition for distribution $A$ to dominate distribution $B$ is that of first order stochastic dominance. That is

$$\Delta F = F_A\left(y_j\right) - F_B\left(y_j\right) \leq 0$$

for all $y_j$ and with the strict inequality holding for some $y_j$. When other attributes are included, the second derivative assumptions $U_{12} \leq 0, U_{13} \leq 0, U_{14} \leq 0, U_{23} \leq 0, U_{24} \leq 0, U_{34} \leq 0,$ imply that each attribute is substitutable for another. For example, $U_{12} \leq 0$ means that the marginal increase in well-being associated with an increase in individual disposable income is decreasing with an increasing level of health. The assumptions also imply that each attribute can be used to compensate insufficiency in other well-being attributes. The next-to-last line in $\mathbf{U}^1$, $U_{123} \geq 0,$ assumes that the person with the highest claim to national aid for compensating for his/her mental health condition is the least educated one. The conditions $U_{124} \geq 0, U_{134} \geq 0, U_{234} \geq 0$ can be interpreted in a similar way. The last line in $\mathbf{U}^1$, $U_{1234} \leq 0$, assumes that the marginal increase in well-being of a person with income support that compensates for mental health and education is decreasing with an increasing level of happiness.

In Section E of the supplementary material, we prove that a condition for the welfare in $A$ to be no lower than in $B$ for all $U \in \mathbf{U}^1$ is

$$\Delta F(y_1, y_2, y_3, y_4) = F_A(y_1, y_2, y_3, y_4) - F_B(y_1, y_2, y_3, y_4) \leq 0$$

for all $y_1, y_2, y_3$ and $y_4$. If this condition holds the first order stochastic dominance conditions hold for each of the attributes. That is, $\Delta F_1(y_1) \leq 0$ for all $y_1$, $\Delta F_2(y_2) \leq 0$ for all $y_2$, $\Delta F_3(y_3) \leq 0$ for all $y_3$, and $\Delta F_4(y_4) \leq 0$ for all $y_4$. We calculate the posterior probabilities for each of these first order stochastic dominance conditions, relevant for a utility function with one attribute, as well as the posterior probability that $\Delta F(y_1, y_2, y_3, y_4) \leq 0$ holds for all $(y_1, y_2, y_3, y_4)$, the condition when all attributes are included. For computing the posterior probability for the complete class of $\mathbf{U}^1$ functions, $99 \times 99 \times 5 \times 5$ function comparisons need to be made for each MCMC draw. A total of 10,000 MCMC draws were used, and posterior probabilities were computed for comparison between years (2001, 2010), (2010, 2015) and (2015, 2019).

The first-order and second-order stochastic dominance results for each of the attributes separately (denoted by FSD and SSD, respectively), along with the changes in the posterior means of the means of each attribute are presented in Table 3. Here, we are concerned with FSD; discussion of SSD is considered in conjunction with the $\mathbf{U}^3$ class of utility functions. Considering each of the attributes in turn, we find that, for income, there is strong evidence that 2010 FSD 2001, but little evidence of dominance in the following two time intervals despite the relatively large change in the mean from 2015 to 2019. For health and happiness no dominance has the highest probability in all time intervals. There is a moderate probability (0.39) that health deteriorated from 2010 to 2015, consistent with a decline in the mean. There are small FSD probabilities (0.28 and 0.35) that happiness improved in the later two time intervals, outcomes that are also consistent with the changes in the means. For education, there is strong evidence of improvement in each of the time intervals with posterior probabilities of FSD of 1.00, 1.00 and 0.87.

The vastly different results for each of the attributes, particularly for health and education, suggest that dominance for the joint distributions of all attributes would be difficult to obtain. That was indeed the outcome. The proportion of MCMC draws that satisfied the conditions for all $(y_1, y_2, y_3, y_4)$ combinations was zero for each of the pairwise comparisons (2001, 2010), (2010, 2015) and (2015, 2019).

Table 3. FSD and SSD Probabilities and Changes in the Means for the Marginal Distributions of Income, Mental Health, Education Level and Happiness

| Criterion | Income | Health | Education | Happiness |
|---|---|---|---|---|
| | | $A = 2001,\ B = 2010$ | | |
| $A$ FSD $B$ | 0.0000 | 0.0000 | 0.0000 | 0.0006 |
| $B$ FSD $A$ | 0.9961 | 0.0017 | 0.9997 | 0.0000 |
| no dominance | 0.0039 | 0.9983 | 0.0003 | 0.9994 |
| $\bar{y}_j(B) - \bar{y}_j(A)$ | 85.81 | 0.0030 | 0.2419 | –0.0382 |
| $A$ SSD $B$ | 0.0000 | 0.0000 | 0.0000 | 0.0012 |
| $B$ SSD $A$ | 0.9998 | 0.7273 | 1.0000 | 0.0112 |
| no dominance | 0.0002 | 0.2727 | 0.0000 | 0.9876 |
| | | $A = 2010,\ B = 2015$ | | |
| $A$ FSD $B$ | 0.0002 | 0.3941 | 0.0000 | 0.0017 |
| $B$ FSD $A$ | 0.3452 | 0.0000 | 1.0000 | 0.2818 |
| no dominance | 0.6546 | 0.6059 | 0.0000 | 0.7165 |
| $\bar{y}_j(B) - \bar{y}_j(A)$ | 8.9 | –0.0057 | 0.2017 | 0.0279 |
| $A$ SSD $B$ | 0.0011 | 0.9010 | 0.0000 | 0.0242 |
| $B$ SSD $A$ | 0.7558 | 0.0000 | 1.0000 | 0.3564 |
| no dominance | 0.2431 | 0.0990 | 0.0000 | 0.6194 |
| | | $A = 2015,\ B = 2019$ | | |
| $A$ FSD $B$ | 0.0000 | 0.0000 | 0.0000 | 0.0032 |
| $B$ FSD $A$ | 0.1621 | 0.0000 | 0.8712 | 0.3467 |
| no dominance | 0.8379 | 1.0000 | 0.1288 | 0.6501 |
| $\bar{y}_j(B) - \bar{y}_j(A)$ | 20.7 | –0.0130 | 0.0937 | 0.0178 |
| $A$ SSD $B$ | 0.0000 | 0.0000 | 0.0000 | 0.0136 |
| $B$ SSD $A$ | 0.2100 | 0.0000 | 0.9999 | 0.6010 |
| no dominance | 0.7900 | 1.0000 | 0.0001 | 0.3854 |

*Notes*: $\bar{y}_j(B) - \bar{y}_j(A)$ refers to the difference of the earlier-year posterior mean from the later-year posterior mean for each of the attributes.

### 5.3 Class of Utility Functions $\mathbf{U}^2$

The $\mathbf{U}^1$ assumption that a high level in one attribute can be used to compensate insufficiency in other well-being attributes is questionable, particularly with respect to mental health and education where $U_{23} \leq 0$. Muller and Trannoy (2011) argue that there is not enough motivation to impose the negative cross partial derivative between education and health in the context of the distribution of international aid. The same applies to the cross partial derivatives between health and happiness and education level and happiness. In the second class of well-being functions $\mathbf{U}^2$, we relax these cross partial derivative conditions by setting $U_{23} = 0$, $U_{24} = 0$, and $U_{34} = 0$. Doing so implies that the well-being function is additively separable with respect to the education level, mental health, and happiness dimensions. It also assumes that mental health and education, mental health and happiness, and education level and happiness are neither substitutes nor complement in well-being.

The resulting class of well-being functions $\mathbf{U}^2$ is

$$\mathbf{U}^2 = \left\{ \begin{array}{c} U_1, U_2, U_3, U_4 \geq 0 \\ U_{12} \leq 0, U_{13} \leq 0, U_{14} \leq 0, U_{23} = 0, U_{24} = 0, U_{34} = 0 \\ U_{123} = 0, U_{124} = 0, U_{134} = 0, U_{234} = 0 \\ U_{1234} = 0 \end{array} \right\}$$

A condition for the level of welfare in $A$ to be no lower than in $B$ for all utility functions $U \in \mathbf{U}^2$ is that $\Delta F(y_1, y_2) \leq 0$ for all $y_1$ and $y_2$, $\Delta F(y_1, y_3) \leq 0$ for all $y_1$ and $y_3$, and $\Delta F(y_1, y_4) \leq 0$ for all $y_1$ and $y_4$. The proof is given in Section E of the supplementary material. Note that the condition $\Delta F(y_1, y_2) \leq 0$ for all $y_1$ and $y_2$ refers to the bivariate first-order dominance condition over the supports of income and mental health only, $\Delta F(y_1, y_3) \leq 0$ for all $y_1$ and $y_3$ refers to the bivariate first-order dominance condition over the supports of income and education only, and $\Delta F(y_1, y_4) \leq 0$ for all $y_1$ and $y_4$ refers to the bivariate first-order dominance condition over the supports of income and happiness only.

For comparison with the bivariate first-order stochastic dominance results we compute bivariate welfare indices for income with each of the other attributes; for all attributes we compute a four-dimensional MWI. These indices are based on a weighted average of well-being attributes like the Human Development Index. Each of the well-being attributes is normalized such that they all lie between 0 and 1. For the years of education and happiness scores, the normalized variable $y_{ij}^{n}$ used to calculate the index is defined as

$$y_{ij}^{n}=\frac{y_{ij}-y_{j}^{m}}{y_{j}^{M}-y_{j}^{m}} \qquad j=3,4$$

where $y_{j}^{m}$ is the minimum benchmark and $y_{j}^{M}$ is the maximum benchmark for well-being indicators $j=3,4$. The normalized variable for income is

$$y_{i1}^{n}=\frac{\log y_{i1}-\log y_{i1}^{m}}{\log y_{i1}^{M}-\log y_{i1}^{m}},$$

for $i=1,2,\ldots,N$, where $N$ is sample size. No normalization is needed for the mental health scores since they are defined in the [0,1] interval. The $MWI$ is a simple average of the normalized attributes

$$MWI=\frac{1}{N}\sum_{i=1}^{N}\frac{1}{L}\left(\sum_{j=1}^{L}y_{ij}^{n}\right)$$

where $L=2$ for a bivariate index, $L=4$ for a four-dimensional index.

In the second column of Table 4, we report changes in the bivariate indices for income with each of the other attributes for the periods (2001, 2010), (2010, 2015) and (2015, 2019). In the fourth column are the corresponding bivariate FSD probabilities relevant for utility functions comprising two attributes, namely, $U(y_1,y_2)$, $U(y_1,y_3)$ and $U(y_1,y_4)$. The only strong evidence of bivariate dominance is for income and education from 2001 to 2010 where we find $\Pr(2010 \text{ FSD } 2001)=0.996$. The corresponding probabilities for the later time intervals are much smaller, specifically $\Pr(2015 \text{ FSD } 2010) = 0.3446$ and $\Pr(2019 \text{ FSD } 2015) = 0.1366$. For the other two bivariate distributions, there is strong evidence of no dominance in all three time intervals. The bivariate improvement in income and education is supported by the index changes $MWI(B)-MWI(A)$. These changes are larger than those for any of the other pairs of attributes. The indices for these other pairs show a deterioration in welfare from 2001 to 2010 and again from 2010 to 2015, and an improvement from 2015 to 2019. However, when all four attributes are included to calculate $MWI$, the changes are all positive; the results (not included in Table 4) are 0.0035 from 2001 to 2010, 0.0112 from 2010 to 2015 and 0.0071 from 2015 to 2019.

Table 4 Welfare index changes and $\mathbf{U}^2$ and $\mathbf{U}^3$ bivariate stochastic dominance probabilities

| Attributes | *MWI*(*B*) – *MWI*(*A*) | Dominance outcome | FSD $\Delta F(y_i, y_j) \le 0$ | $\Delta P_{1j}(z_1; y_j) \le 0$ | SSD $\Delta P_{1j}(z_1; y_j) \le 0$ $\Delta H_j(y_j) \le 0$ |
|---|---|---|---|---|---|
| | | | $A = 2001,\ B = 2010$ | | |
| Income and health | –0.0183 | A dom B | 0.0000 | 0.0000 | 0.0000 |
| | | B dom A | 0.0136 | 0.7546 | 0.6635 |
| | | No dom | 0.9864 | 0.2454 | 0.3365 |
| Income and education | 0.0104 | A dom B | 0.0000 | 0.0000 | 0.0000 |
| | | B dom A | 0.9960 | 0.9998 | 0.9998 |
| | | No dom | 0.0040 | 0.0002 | 0.0002 |
| Income and happiness | –0.0246 | A dom B | 0.0000 | 0.0000 | 0.0000 |
| | | B dom A | 0.0000 | 0.0748 | 0.0089 |
| | | No dom | 1.0000 | 0.9252 | 0.9252 |
| | | | $A = 2010,\ B = 2015$ | | |
| Income and health | –0.0064 | A dom B | 0.0000 | 0.0004 | 0.0004 |
| | | B dom A | 0.0000 | 0.0000 | 0.0000 |
| | | No dom | 1.0000 | 0.9996 | 0.9996 |
| Income and education | 0.0216 | A dom B | 0.0000 | 0.0000 | 0.0000 |
| | | B dom A | 0.3446 | 0.7545 | 0.7545 |
| | | No dom | 0.6554 | 0.2455 | 0.2455 |
| Income and happiness | –0.0001 | A dom B | 0.0000 | 0.0000 | 0.0000 |
| | | B dom A | 0.0311 | 0.0946 | 0.0892 |
| | | No dom | 0.9689 | 0.9054 | 0.9108 |
| | | | $A = 2015,\ B = 2019$ | | |
| Income and health | 0.0003 | A dom B | 0.0000 | 0.0000 | 0.0000 |
| | | B dom A | 0.0000 | 0.0000 | 0.0000 |
| | | No dom | 1.0000 | 1.0000 | 1.0000 |
| Income and education | 0.0186 | A dom B | 0.0000 | 0.0000 | 0.0000 |
| | | B dom A | 0.1366 | 0.1903 | 0.1903 |
| | | No dom | 0.8634 | 0.8097 | 0.8097 |
| Income and happiness | 0.0091 | A dom B | 0.0000 | 0.0000 | 0.0000 |
| | | B dom A | 0.0192 | 0.0401 | 0.0392 |
| | | No dom | 0.9808 | 0.9599 | 0.9608 |

To obtain a nonzero probability of dominance under utility function class $\mathbf{U}^2$ with all four attributes, the proportion of MCMC draws that satisfy all three bivariate FSD conditions simultaneously must be nonzero. As one might expect from the lack of dominance for the income-health and income-happiness conditions, this proportion of MCMC draws was zero, leading to a probability of one of no dominance for each of the time intervals.

### 5.4 Class of Utility Functions $\mathbf{U}^3$

Even after setting the conditions $U_{23} = 0, U_{24} = 0,$ and $U_{34} = 0,$ it is still highly unlikely that one can generate unambiguous rankings when comparing well-being distributions without adding further restrictions on the utility function. In the third class of well-being functions, the following conditions are added: $U_{11}, U_{22}, U_{33}, U_{44} \leq 0.$ These conditions imply that there is preference for more equal marginal distributions for each well-being attribute. They introduce some concern for health, education, and happiness inequalities. Further restrictions are also introduced: $U_{121} \geq 0,$ $U_{131} \geq 0,$ and $U_{141} \geq 0.$ The first of these conditions can be thought of as a condition that the person with the highest claim for national aid for compensating a low mental health score is the poorest one; $U_{131} \geq 0,$ and $U_{141} \geq 0$ can be thought of in a similar way with respect to the education level and happiness variables. Providing better access to mental health care and education for income-poor people are examples of policy applications.

The third class of well-being functions $\mathbf{U}^3$ is

$$\mathbf{U}^3 = \left\{ \begin{array}{c} U_1, U_2, U_3, U_4 \geq 0 \\ U_{12} \leq 0, U_{13} \leq 0, U_{14} \leq 0, U_{23} = 0, U_{24} = 0, U_{34} = 0, \\ U_{11} \leq 0, U_{22} \leq 0, U_{33} \leq 0, U_{44} \leq 0, \\ U_{121} \geq 0, U_{131} \geq 0, U_{141} \geq 0, \\ U_{123} = 0, U_{124} = 0, U_{134} = 0, U_{234} = 0, \\ U_{1234} = 0 \end{array} \right\}$$

Before defining a set of dominance conditions for $\mathbf{U}^3$, we define the following notation relevant for second order stochastic dominance relationships:

$$H_i(y_i) = \int_0^{y_i} F_i(s)ds$$

$$H_i(y_i; y_j, y_k, y_l) = \int_0^{y_i} F(r, y_j, y_k, y_l)\, dr$$

and

$$H_i(y_i;y_j) = H_i(y_i;y_j,a_k,a_l)$$

for any $i,j,k,l$, where $F_1,F_2,F_3$ and $F_4$ are the marginal distribution functions of income, mental health score, education, and happiness, respectively; $a_k$ and $a_l$ are the upper limits of $Y_k$ and $Y_l$, respectively. For the class of utility functions, $\mathbf{U}^3$, a set of stochastic dominance conditions for the level of welfare in $A$ to be no lower than in $B$ for all utility functions $U \in \mathbf{U}^3$ is

1. $\Delta H_j\left(y_j\right) \leq 0,$ for all $y_j$, $j = 2,3,4,$ where $\Delta H = H_A(\boldsymbol{y}) - H_B(\boldsymbol{y}).$
2. $\Delta H_1\left(y_1;y_2\right) \leq 0,$ for all $y_1$ and $y_2$,
3. $\Delta H_1\left(y_1;y_3\right) \leq 0,$ for all $y_1$ and $y_3$,
4. $\Delta H_1\left(y_1;y_4\right) \leq 0,$ for all $y_1$ and $y_4$.

The proof is given in Section E of the online supplementary material.

The condition $\Delta H_j\left(y_j\right) \leq 0$ for all $y_j$ is the condition for $F_j$ at time $A$ to second-order stochastically dominate $F_j$ at time $B$. If utility is a function of a single attribute, $U\left(y_j\right)$, then it is the relevant condition for utility functions with the properties $U_j \geq 0$ and $U_{jj} \leq 0$. For utility functions with two attributes, say $y_1$ and $y_j$, the SSD conditions relevant for class $\mathbf{U}^3$ are $\Delta H_j\left(y_j\right) \leq 0$ for all $y_j$ and $\Delta H_1\left(y_1;y_j\right) \leq 0$ for all $y_1$ and $y_j$.

Since $Y_3$ and $Y_4$ are categorical variables, it is straightforward to check whether the conditions $\Delta H_3\left(y_3\right) \leq 0$ and $\Delta H_4\left(y_4\right) \leq 0$ hold. A more convenient condition than the remaining dominance conditions defined in 1 – 4 above can be obtained from the equivalent poverty ordering. Foster and Shorrocks (1988) and Duclos et al. (2006) showed that poverty gap conditions are equivalent to second order stochastic dominance. Given a poverty line for mental health $z_2$, the poverty gap is

$$P_2(z_2) = \int_0^{z_2} \left(z_2 - y_2\right) p_2\left(y_2\right) dy_2$$

The condition $\Delta H_2\left(y_2\right)\leq 0$ for all $y_2$ is equivalent to the poverty gap in $A$ being no larger than the poverty gap in $B$ for all poverty lines $z_2$. For the conditions involving income $y_1$, we define the poverty gaps $P_{1j}\left(z_1;y_j\right), j=2,3,4$ as

$$P_{1j}\left(z_1;y_j\right)=\int_0^{y_j}\int_0^{z_1}\left(z_1-y_1\right)p\left(y_1,y_j\right)dy_1\,dy_j$$

$P_{12}\left(z_1;y_2\right)$ is the absolute poverty gap for a compensating variable (income) for a population below an income level of $z_1$ and below a mental health level of $Y_2$. Similar interpretations hold for $P_{1j}\left(z_1;y_j\right), j=3,4$. Assessing changes in $H_2(y_2)$, $H_1(y_1;y_2)$, $H_1(y_1;y_3)$, and $H_1(y_1;y_4)$ is equivalent to assessing changes in the poverty gap conditions, $P_2(z_2)$, $P_{12}(z_1;y_2)$, $P_{13}(z_1;y_3)$ and $P_{14}(z_1;y_4)$, where $z_1$ and $z_2$ are poverty lines for income and mental health, respectively. The condition $\Delta H_1\left(y_1;y_j\right)\leq 0$ for all $y_1$ and $y_j$ requires that $P_{1j}\left(z_1;y_j\right)$ is no larger in $A$ than in $B$ for all values of $z_1$ and $y_j$, for $j=2,3,4$.

The policy implication of conditions for utility functions belonging to the class $\mathbf{U}^3$ is that improving mental health, happiness, and the education system could be pursued by the government or policy makers by cash transfer policies directed to the mentally ill, the unhappy, and uneducated persons. Note that the conditions $\Delta P_2(z_2)\leq 0$ and $\Delta P_{12}(z_1;y_2)\leq 0$ for all $y_1, y_2$ refer to the dominance condition of $U\in\mathbf{U}^3$ over the space of income and mental health only, $\Delta H_3(y_3)\leq 0$ and $\Delta P_{1,3}(z_1;y_3)\leq 0$ for all $y_1$, $y_3$ refers to the dominance condition of $U\in\mathbf{U}^3$ over the space of income and education only, and $\Delta H_4(y_4)\leq 0$ and $\Delta P_{1,4}(z_1;y_4)\leq 0$ for all $y_1$, $y_4$ refers to the dominance condition of $U\in\mathbf{U}^3$ over the space of income and happiness only.

As was the case with classes of utility functions $\mathbf{U}^1$ and $\mathbf{U}^2$, when all attributes were included, it proved difficult to establish any evidence of dominance for the three pairs of years and the utility class $\mathbf{U}^3$. In one case, there was a very small number of MCMC draws that did satisfy the conditions, leading to a probability of dominance of $\Pr\left(2010\ SD(\mathbf{U}^3)\ 2001\right)=0.0084$, a value not large enough to challenge a conclusion of no dominance. There were, however, some more definitive results when we consider utility functions with only one attribute or only two attributes, income and one of the others. For only one attribute, the relevant results are the SSD probabilities in Table 3. Restricting the class of utility functions such that they are concave $\left(U_{jj} \leq 0\right)$ as well as increasing $\left(U_j \geq 0\right)$ has meant that dominance is more easily established in some cases. Specifically, for income we find $\Pr\left(2015 \text{ FSD } 2010\right) = 0.3452$ has increased to $\Pr\left(2015 \text{ SSD } 2010\right) = 0.7558,$ for health $\Pr\left(2010 \text{ FSD } 2001\right) = 0.0017$ has increased to $\Pr\left(2010 \text{ SSD } 2001\right) = 0.7273,$ and, for happiness, $\Pr\left(2019 \text{ FSD } 2015\right) = 0.346$ has increased to $\Pr\left(2019 \text{ SSD } 2015\right) = 0.6010.$

The results for bivariate dominance are presented in the last two columns of Table 4. Changes in the bivariate poverty gaps for income and each of the other attributes are considered with and without the extra second order stochastic dominance constraint $\Delta H_j(y_j) \leq 0,\ j = 2,3,4.$ Both $\Delta P_{1j}(z_1; y_j) \leq 0$ and $\Delta H_j(y_j) \leq 0$ comprise the bivariate conditions for $\mathbf{U}^3$. For this class of utility functions, we are not only interested in knowing whether income has played some unambiguous role in compensating for health, education and happiness, but also the effectiveness of policy that gives a priority to poor/disadvantaged people in distributing national aid to improve well-being in terms of mental health, the level of education, and happiness.

Comparing 2001 and 2010, we find the 2010 absolute poverty gaps $P_{12}(z_1; y_2)$ and $P_{13}(z_1; y_3)$ are smaller than those of 2001 with probabilities 0.7546 and 0.9998, respectively, but there is only a small probability that $P_{14}(z_1; y_4)$ in 2010 is smaller than in 2001. Also, there is some evidence that 2010 dominates 2001 for the class of utility function $U \in \mathbf{U}^3$ in the space of income and mental health and income and education, but not for income and happiness. These results imply that income has played a significant role in compensating for the lack of mental health and education separately, but not happiness.

For the other two pairs of years, the only change that yields evidence of dominance is that for 2015 dominating 2010 for income and education where the probability of dominance for both the poverty gap and $\mathbf{U}^3$ is 0.75.

# 6. POVERTY

As well as being interested in whether well-being has improved over time, we are also likely to be concerned with whether poverty has been reduced. To examine this question we consider both single index measures and a restricted form of stochastic dominance.

## 6.1 Single Index Measures

The most common single index measure, and the one that we consider here, is the headcount ratio. In a single dimension, this ratio is simply the proportion of individuals whose value of the single attribute falls below a pre-defined poverty line. We will denote this proportion as $HC_1$. When we move to more than one dimension, the headcount concept can be extended in a variety of ways. With two attributes, one way is to count the proportion of individuals whose attribute values fall below both poverty lines. We will denote this proportion as $HC_2$. Another way is to regard an individual as poor if they lie below the poverty line in one or both of the two dimensions. For more than two dimensions, these options have been placed within a general framework developed by Alkire and Foster (2011). In this framework, finding a multidimensional headcount involves two cut-offs. The first cutoff is a dimension-specific deprivation cutoff which identifies whether a person is deprived with respect to a particular dimension. The second cut-off refers to the minimum number of deprivations that need to be experienced by an individual before they are regarded as multidimensionally poor. In our study, where there are four attributes, there are four possible cut-offs each of which define a different multidimensional poverty index. We denote these indexes as $MH_j \;\; j = 1,2,3,4$ where $j$ is the minimum number of deprivations needed for an individual to be considered multidimensionally poor. One drawback with this measure is that if a poor person becomes deprived in a new dimension, then $MH_j$ remains unchanged. To overcome this problem an alternative measure which adjusts $MH_j$ by multiplying it by the average deprivation share across the poor has been suggested. This adjustment has the effect of replacing the number of poor individuals with the average number of deprivations experienced by each of the poor. Specifically, the new measure is given by

$$MHA_j = MH_j \times A_j$$
$$= \frac{\text{total number of deprivations exprienced by the poor using cutoff } j}{J \times N}$$

where $A_j$ is the average deprivation share across the poor and $J$ is the number of attributes (4 in our case).[2]

### 6.2 Restricted Stochastic Dominance

Another way to assess whether poverty has been reduced is via a restricted form of stochastic dominance where the left tail of the distributions from different time periods are compared. If we are considering a single attribute, then first and second-order stochastic dominance, relevant under abbreviated forms of $\mathbf{U}^1$ and $\mathbf{U}^3$, can be assessed by counting the proportion of MCMC parameter draws that satisfy the dominance constraints for all values of the attribute up to the poverty line. This proportion represents the posterior probability that poverty has been reduced according to the stochastic dominance criterion. Similarly, bivariate and multivariate conditions can be assessed under the assumptions pertaining to utility function classes $\mathbf{U}^1, \mathbf{U}^2$ and $\mathbf{U}^3$. For example, for the class of well-being function $\mathbf{U}^3$, we can use the poverty gap conditions defined above up to some fixed poverty lines for income, mental health, education, and happiness.

### 6.3 Results

We consider results for each attribute separately (Table 5), results from bivariate distributions of income coupled with each of the other attributes (Table 6), and then multidimensional results from all four attributes (Table 7). The poverty lines used for each of the attributes were \$20,164 for income[3], 0.5 for mental health, and category 2 for both education and happiness (see Tables 1 and 2).

For income in Table 5, both the stochastic dominance results and the headcount changes show a decline in poverty from 2001 to 2010 and again from 2010 to 2015. There is less evidence of improvement from 2015 to 2019, however. In this time interval, the decline in the headcount ratio is less pronounced and the highest dominance probability for both FSD and SSD is that of no dominance.

[2] Precise definitions useful for computing these values are provided in Section G of the supplementary material.

[3] This value was the grid point closest to \$20,000.

For mental health, there is evidence of improvement from 2001 to 2010, but then a decline from 2010 to 2015. In the last time interval, there is no evidence of dominance but the proportion of individuals with poor mental health has increased. Education shows a strong improvement in all three time periods. The proportion of unhappy people declines in each of the time periods. This decline is supported by some evidence of the later years stochastically dominating the earlier years.

Table 5. Restricted FSD and SSD Posterior Probabilities for Poverty Improvement and Headcount Changes for the Marginal Distributions of Income, Mental Health, Education Level and Happiness

| Criterion | Income | Health | Education | Happiness |
|---|---|---|---|---|
| | | $A = 2001,\ B = 2010$ | | |
| $A$ FSD $B$ | 0.0000 | 0.0000 | 0.0000 | 0.0052 |
| $B$ FSD $A$ | 0.9997 | 0.5198 | 1.0000 | 0.9353 |
| no dominance | 0.0003 | 0.4802 | 0.0000 | 0.0595 |
| $HC_1(A) - HC_1(B)$ | 0.1264 | 0.0009 | 0.0933 | 0.0057 |
| $A$ SSD $B$ | 0.0000 | 0.0000 | 0.0000 | 0.0068 |
| $B$ SSD $A$ | 0.9998 | 0.9302 | 1.0000 | 0.9683 |
| no dominance | 0.0002 | 0.0698 | 0.0000 | 0.0249 |
| | | $A = 2010,\ B = 2015$ | | |
| $A$ FSD $B$ | 0.0019 | 0.7950 | 0.0000 | 0.2046 |
| $B$ FSD $A$ | 0.7814 | 0.0000 | 1.0000 | 0.3995 |
| no dominance | 0.2167 | 0.2050 | 0.0000 | 0.3959 |
| $HC_1(A) - HC_1(B)$ | 0.0181 | -0.0064 | 0.0658 | 0.0020 |
| $A$ SSD $B$ | 0.0065 | 0.9710 | 0.0000 | 0.2831 |
| $B$ SSD $A$ | 0.7926 | 0.0000 | 1.0000 | 0.4188 |
| no dominance | 0.2009 | 0.0290 | 0.0000 | 0.2981 |
| | | $A = 2015,\ B = 2019$ | | |
| $A$ FSD $B$ | 0.0630 | 0.0000 | 0.0000 | 0.1191 |
| $B$ FSD $A$ | 0.1883 | 0.0000 | 0.9980 | 0.4992 |
| no dominance | 0.7487 | 1.0000 | 0.0030 | 0.3812 |
| $HC_1(A) - HC_1(B)$ | 0.0094 | -0.0217 | 0.0277 | 0.0003 |
| $A$ SSD $B$ | 0.2580 | 0.0000 | 0.0000 | 0.1299 |
| $B$ SSD $A$ | 0.2100 | 0.0023 | 1.0000 | 0.6318 |
| no dominance | 0.5320 | 0.9977 | 0.0000 | 0.2383 |

Table 6 $\mathbf{U}^2$ and $\mathbf{U}^3$ Bivariate Restricted Stochastic Dominance Posterior Probabilities for Poverty Improvement Under $\mathbf{U}^2$ and $\mathbf{U}^3$ and Changes in Bivariate Headcounts

| Attributes | $HC_2(A)$ – $HC_2(B)$ | Dominance outcome | FSD $\Delta F(y_i, y_j) \le 0$ | $\Delta P_{1j}(z_1; y_j) \le 0$ | SSD $\Delta P_{1j}(z_1; y_j) \le 0$ $\Delta H_j(y_j) \le 0$ |
|---|---|---|---|---|---|
| | | | $A = 2001,\ B = 2010$ | | |
| Income and health | 0.0139 | A dom B | 0.0000 | 0.0000 | 0.0000 |
| | | B dom A | 0.8912 | 0.9079 | 0.8567 |
| | | No dom | 0.1088 | 0.0921 | 0.1433 |
| Income and education | 0.0953 | A dom B | 0.0000 | 0.0000 | 0.0000 |
| | | B dom A | 0.9999 | 0.9999 | 0.9999 |
| | | No dom | 0.0001 | 0.0001 | 0.0001 |
| Income and happiness | 0.0053 | A dom B | 0.0000 | 0.0000 | 0.0000 |
| | | B dom A | 0.9505 | 0.9605 | 0.9350 |
| | | No dom | 0.0495 | 0.0395 | 0.0650 |
| | | | $A = 2010,\ B = 2015$ | | |
| Income and health | 0.0019 | A dom B | 0.0026 | 0.0433 | 0.0433 |
| | | B dom A | 0.0000 | 0.0000 | 0.0000 |
| | | No dom | 0.9974 | 0.9567 | 0.9567 |
| Income and education | 0.0224 | A dom B | 0.0000 | 0.0000 | 0.0000 |
| | | B dom A | 0.8806 | 0.8881 | 0.8881 |
| | | No dom | 0.1194 | 0.1119 | 0.1119 |
| Income and happiness | 0.0004 | A dom B | 0.0785 | 0.1348 | 0.0905 |
| | | B dom A | 0.1521 | 0.2207 | 0.1665 |
| | | No dom | 0.7694 | 0.6445 | 0.7430 |
| | | | $A = 2015,\ B = 2019$ | | |
| Income and health | -0.0027 | A dom B | 0.0000 | 0.0000 | 0.0000 |
| | | B dom A | 0.0011 | 0.0032 | 0.0002 |
| | | No dom | 0.9989 | 0.9968 | 0.9998 |
| Income and education | 0.0082 | A dom B | 0.0072 | 0.0688 | 0.0000 |
| | | B dom A | 0.2273 | 0.2507 | 0.2507 |
| | | No dom | 0.7655 | 0.6805 | 0.7493 |
| Income and happiness | 0.0001 | A dom B | 0.0875 | 0.1695 | 0.0724 |
| | | B dom A | 0.0983 | 0.1380 | 0.1249 |
| | | No dom | 0.8142 | 0.6925 | 0.8027 |

Examining the bivariate results presented in Table 6 for income with each of the other attributes, we find positive improvement for each of the bivariate distributions from 2001 to 2010, results which are consistent with those from considering each attribute separately. From 2010 to 2015, no dominance is the most likely outcome for income and mental health, reflecting the contrasting results for each of these attributes when they are considered separately. To a lesser extent the same is true for income and happiness; the income and education distribution shows strong evidence of dominance. There is a reduction in all three headcounts, although the change is small for income/health and income/happiness. In the third period from 2015 to 2019, no dominance has the highest posterior probability for all bivariate distributions and changes in the headcounts are small.

Results in Table 7, obtained by an all four attributes are considered jointly, reflect those from the univariate and bivariate results. Changes in the multivariate poverty indexes are highest for the period 2001 to 2010 and this is the only period where there is evidence of the later year dominating the earlier year.

Table 7 Four-Dimensional Restricted Stochastic Dominance Posterior Probabilities for Poverty Improvement Under $\mathbf{U}^1$, $\mathbf{U}^2$ and $\mathbf{U}^3$, and Changes in Multidimensional Poverty Measures

| Criterion | $A=2001$, $B=2010$ | $A=2010$, $B=2015$ | $A=2015$, $B=2019$ |
|---|---|---|---|
| $MH_1(A)-MH_1(B)$ | 0.1038 | 0.0515 | 0.0162 |
| $MH_2(A)-MH_2(B)$ | 0.0976 | 0.0241 | 0.0012 |
| $MH_3(A)-MH_3(B)$ | 0.0136 | 0.0037 | –0.0016 |
| $MH_4(A)-MH_4(B)$ | 0.0014 | 0.0003 | –0.0001 |
| $MHA_1(A)-MHA_1(B)$ | 0.0541 | 0.0199 | 0.0039 |
| $MHA_2(A)-MHA_2(B)$ | 0.0525 | 0.0130 | 0.0002 |
| $MHA_3(A)-MHA_3(B)$ | 0.0105 | 0.0029 | –0.0012 |
| $MHA_4(A)-MHA_4(B)$ | 0.0014 | 0.0003 | –0.0001 |
| $A$ dom $B$ ($U^1$) | 0.0000 | 0.0026 | 0.0000 |
| $A$ dom $B$ ($U^2$) | 0.0000 | 0.0000 | 0.0000 |
| $A$ dom $B$ ($U^3$) | 0.0000 | 0.0000 | 0.0000 |
| $B$ dom $A$ ($U^1$) | 0.3347 | 0.0000 | 0.0022 |
| $B$ dom $A$ ($U^2$) | 0.8574 | 0.0000 | 0.0005 |
| $B$ dom $A$ ($U^3$) | 0.8108 | 0.0000 | 0.0001 |
| No dom ($U^1$) | 0.6653 | 0.9974 | 0.9978 |
| No dom ($U^2$) | 0.1426 | 1.0000 | 0.9995 |
| No dom ($U^3$) | 0.1892 | 1.0000 | 0.9999 |

# 7. CONCLUSIONS

Using four important well-being attributes, income, education, mental health, and happiness, we have illustrated how flexible multivariate Bayesian techniques can be used to assess whether improvements in welfare have been realized over time. Flexible mixture distributions are used to model the continuous variables income and mental health scores, while data on education and happiness are in the form of discrete ordinal distributions. The marginal distributions for these variables are conveniently combined into a multivariate distribution using a copula and estimated using Bayesian MCMC techniques. Improvements in both well-being generally and poverty magnitude are assessed using posterior means of single index measures and posterior probabilities of stochastic dominance. These assessments are made for each attribute considered separately, from the bivariate distributions of pairs of attributes, and from the four-dimensional distribution of all attributes. The conditions for stochastic dominance depend on the properties of the assumed class of utility function used to define well-being. Posterior probabilities are found for three classes of utility functions. The techniques are applied to Australian data over the period 2001 to 2019. For well-being generally, they reveal an overall improvement in education, a qualified improvement in income, a decline in mental health after 2010, and a qualified improvement in happiness after 2010. Similar results, but with some slight modifications, were obtained when the analysis was restricted to examination of poverty. There was a more definite reduction in the income dimension of poverty, at least until 2015, mental health improvement until 2010 but declined thereafter, and there was a definite improvement in happiness prior to 2010. The contrasting results for the different attributes meant that stochastic dominance could not be established when comparing four-dimensional distributions; nor could it be established for most of the bivariate distributions using income and the other attributes. Exceptions were the distribution for income and education where there were definite improvements prior to 2015 for both general well-being and poverty and for income and mental health where, in terms of the $\mathbf{U}^3$ class of utility functions, there were similar improvements prior to 2010.

## SUPPLEMENTARY MATERIALS

**Appendices**: Details of the parameter settings for the prior distributions are provided in Appendix A. The MCMC sampling scheme is described in Appendix B. Appendix C contains the procedure and results for choosing the number of components in the mixture models. The marginal predictive densities for each attribute are illustrated in Appendix D. Proofs of multidimensional stochastic dominance conditions are provided in Appendix E. Appendix F contains precise statements of the procedure for computing probabilities of stochastic dominance. Details of the computations for the distribution functions necessary for assessing stochastic dominance and for the single index measures appear in Appendix G.

## ACKNOWLEDGEMENTS

This paper uses unit record data from Household, Income and Labour Dynamics in Australia Survey [HILDA] conducted by the Australian Government Department of Social Services (DSS). The findings and views reported in this paper, however, are those of the authors and should not be attributed to the Australian Government, DSS, or any of DSS' contractors or partners.

## REFERENCES

Alkire, S. and Foster, J. (2011), " Counting and Multidimensional Poverty Measurement" *Journal of Public Economics*, 95, 476–487.

Anderson, G. (1996), "Nonparametric Tests for Stochastic Dominance" *Econometrica*, 64, 1183–1193.

Atkinson, A. and Bourguignon, F. (1982), " The Comparison of Multi-dimensioned Distribution of Economic Status" *Review of Economic Studies*, 49, 183–201.

Bourguignon, F. and Chakravarty, S. R. (2003), " The Measurement of Multidimensional Poverty" *Journal of Economic Inequality*, 1, 25–49.

Butterworth, P. and Crosier, T. (2004), "The Validity of SF-36 in an Australian National Household Survey: Demonstrating the Applicability of the Household Income and Labour Dynamics in Australia (HILDA) Survey to Examination of Health Inequalities" *BMC Public Health*,

4(44), 1–11.

Crawford, I. (2005), “ A Nonparametric Test of Stochastic Dominance in Multivariate Distributions” *Discussion Papers in Economics, DP 12/05, University of Surrey*.

Danaher, P. and Smith, M. (2011), “ Modeling Multivariate Distributions Using a Copula: Applications in Marketing” *Marketing Science*, 30, 4-21.

Davidson, R. and Duclos, J. Y. (2000), “Statistical Inference for Stochastic Dominance and for the Measurement of Poverty and Inequality” *Econometrica*, 68(6), 1435– 1464.

Duclos, J. Y., Sahn, D. E., and Younger, S. D. (2006), “ Robust Multidimensional Poverty Comparisons” *The Economic Journal*, 116(514), 943–968.

Foster, J. E. and Shorrocks, A. F. (1988), “Poverty Orderings and Welfare Dominance” *Social Choice and Welfare*, 5 179-198.

Gunawan, D., Griffiths, W. E., and Chotikapanich, D. (2020a), “Posterior Probabilities for Lorenz and Stochastic Dominance of Australian Income Distributions”, *Economic Record*, 97, 504-524.

Gunawan, D., Panagiotelis, A., Griffiths, W. E., and Chotikapanich, D. (2020b), “Bayesian Weighted Inference from Surveys”, *Australian and New Zealand Journal of Statistics*, 62, 71-94.

Hall, P., and Yatchew, A. (2005), “Unified Approach to Testing Functional Hypotheses in Semiparametric Contexts,” *Journal of Econometrics*, 127, 225-252.

Hofert, M., Machler, M., and McNeil, A. J. (2012), “ Likelihood Inference for Archimedean Copulas in High Dimensions Under Known Margins” *Journal of Multivariate Analysis*, 110, 133–150.

Kolm, S. (1977), “Multidimensional Egalitarianism” *Quarterly Journal of Economics*, 91(1), 1–13.

Lander, D., Gunawan, D., Griffiths, W. E., and Chotikapanich, D. (2020), “Bayesian Assessment of Lorenz and Stochastic Dominance” *Canadian Journal of Economics*, 53(2), 767–799.

Lugo, M. A. (2007), “ Comparing Multidimensional Indices of Inequality: Methods and Applications” *Research in Economic Inequality*, 14, 213–236.

Maasoumi, E. (1986), The Measurement and Decomposition of Multidimensional Inequality” *Econometrica*, 54, 991–997.

Maasoumi, E. and Racine, J. S. (2016), "A Solution to Aggregation and an Application to Multidimensional 'Well-Being' Frontiers" *Journal of Econometrics*, 191(2), 374– 383.

Maasoumi, E. and Zandvakili, S. (1986), "A Class of Generalized Measures of Mobility with Applications" *Economics Letters*, 22, 97–102.

McCaig, B. and Yatchew, A. (2007), "International Welfare Comparisons and Nonparametric Testing of Multivariate Stochastic Dominance" *Journal of Applied Econometrics*, 22, 951–969.

Muller, C. and Trannoy, A. (2011), " A Dominance Approach to the Appraisal of the Distribution of Well-Being Across Countries" *Journal of Public Economics*, 95, 239– 246.

Muller, C. and Trannoy, A. (2012), " Multidimensional Inequality Comparisons: A Compensation Perspective" *Journal of Economic Theory*, 147, 1427–1449.

Nelsen, R. B. (2006), *An Introduction to Copulas*, New York: Springer.

Pitt, M., Chan, D., and Kohn, R. (2006), " Efficient Bayesian Inference for Gaussian Copula Regression Models" *Biometrika*, 93(3), 537–554.

Sila, U. and Dugain, V. (2019), "Income, Wealth, and Earnings Inequality in Australia: Evidence from the HILDA Survey" *OECD Economics Department Working Papers No. 1538, Paris.*

Song, P. (2000), "Multivariate Dispersion Models Generated from Gaussian Copula" *Scandinavian Journal of Statistics*, 27(2), 305-320.

Trivedi, P. and Zimmer, D. (2005), " Copula Modeling: An Introduction for Practitioners" *Foundation and Trends in Econometrics*, 1, 1-111.

Ware, J. E. and Gandek, B. (1998), "Overview of the SF-36 Health Survey and the International Quality of Life Assessment (IQOLA) project" *Journal of Clinical Epidemiology*, 11, 903-912.

Ware, J. E., Snow, K. K., Kolinski, M., and Gandeck, B. (1993), *SF-36 Health Survey Manual and Interpretation Guide*. The Health Institute New England Medical Centre, Boston, MA.

Watson, N. and Wooden, M. 2012. "The HILDA Survey: A Case Study in the Design and Development of a Successful Household Panel Study," *Longitudinal and Life Course Studies*, 3, 369–381.